\documentclass[epjST]{svjour}
\usepackage{graphics}

\def\be{\begin{equation}}
\def\ee{\end{equation}}
\def\bc{\begin{center}}
\def\ec{\end{center}}

\def\phiglass{\phi_{\rm glass}}

\begin{document}

\title{Relaxation functions and dynamical heterogeneities in a model of 
chemical gel interfering with glass transition}

\author{Antonio de Candia \inst{1,}\inst{2,}\inst{3}, Annalisa Fierro  \inst{2}, Raffaele Pastore  \inst{2,}\inst{4}, 
Massimo Pica Ciamarra \inst{2,}\inst{5} and Antonio Coniglio \inst{1,}\inst{2}}

\institute{Dipartimento di Fisica, ``Ettore Pancini'', Universit\`a di Napoli 
``Federico II'', Complesso Universitario di Monte Sant'Angelo, via
Cintia, 80126 Napoli, Italy 
\and 
CNR-SPIN, via Cintia, 80126 Napoli, Italy 
\and 
INFN, Sezione di Napoli, via Cintia, 80126 Napoli, Italy
\and 
UC Simulation Center, University of Cincinnati, and Procter \& Gamble Co., Cincinnati, Ohio 45219, USA
\and
Division of Physics and Applied Physics, School of Physical and Mathematical Sciences,
Nanyang Technological University, Singapore 637371, Singapore}

\abstract{
We investigate the heterogeneous dynamics in a model, where chemical gelation and glass transition interplay,
focusing on the dynamical susceptibility. 
Two independent mechanisms give raise to the correlations, which are manifested in the dynamical susceptibility:
one is related to the presence of permanent clusters, 
while the other is due to the increase of particle crowding as the glass transition is approached.
The superposition of these two mechanisms originates a variety of different behaviours.
We show that these two mechanisms can be unentangled considering the wave vector dependence of the dynamical susceptibility.
}
\maketitle
\section{Introduction}
Gels and glasses are both amorphous solids. Gels are elastic disordered solids
observed at low density in systems of molecules bonded to each other through attractive forces or chemical links.
In chemical gels, the transition from sol to gel has been explained \cite{flory,degennes} in terms of the appearance
of a percolating cluster of monomers linked by bonds, that arrests the dynamics in the limit of 
small wave-vector, $k_{min}=2\pi/L$, with $L$ being the system size.
Experimental measurements  have confirmed this geometrical interpretation. 
Indeed, the chemical sol-gel transition shows the same continuous nature of the random percolation transition.
Recently, it has been shown that the same cluster mechanism holds generally for gelling systems \cite{jcp} and 
Mode Coupling Theory (MCT) schematic model A \cite{gotze3}.  
In particular, in Ref.  \cite{arenzon}
scaling predictions for the time correlation function were obtained, and successfully tested in the
$F_{12}$ MCT schematic model and facilitated spin systems on Bethe lattice.
Unlike gels, glasses usually exhibit a structural arrest at high density, 
with the glass transition occurring  also in systems of particles only interacting with excluded volume. 
In this case, the dynamic arrest occurs at all wave-vectors ranging form $k_{min}$ 
to $k_{max}=2\pi/\sigma$, where $\sigma$ is the particle size.
Moreover, the glass transition has been associated to an ideal mixed order transition, where a discontinuous order parameter
is accompanied by a diverging response, while
MCT \cite{gotze3,gotze1,gotze2} well describes dynamical behaviour of glassy systems (if not in the deeply supercooled regime).

Despite these fundamentals differences, it is not always easy to distinguish between gels and glasses. 
To this aim it is useful to consider the presence of Dynamic Heterogeneities (DHs), groups of
particles dynamically correlated over a time scale of the order of the relaxation time.
In particular, the Dynamic Susceptibility $\chi_4(k,t)$, commonly used to measure DHs,
shows a different behaviour on approaching the two transitions \cite{JSTAT_DH}.
For chemical gels, it was  theoretically shown and numerically verified that  
at small wave vector (i.e. $k\rightarrow 0$) and long time (i.e. $t\rightarrow \infty$),
$\chi_4(k,t)$ tends to the mean cluster size, which diverges at the gelation threshold with the exponent $\gamma$ of the random percolation \cite{abete}.
Indeed, in chemical gels, DHs have a clear static origin. 
This is different from what occurs in glassy systems, where  the dynamical susceptibility
displays a maximum in time, whose value increases as the glass transition is approached \cite{het1,het2,het3,het4,het5,het7,het8}.
However, the distinction between gels and glasses may be still elusive when the two transitions coexist, 
as in some polymer or colloidal systems, where a crossover 
from gel-like to glass-like behaviour is observed on varying the control parameters (see, for instance, the  
$PL64/D_2O$ $AHS$ micellar system studied in \cite{chen}).
In this case, the relaxation functions may exhibit complex decays,
such as multi-step and logarithmic decays  \cite{berthier,nagi}. 

In this paper, we consider a model for polymer suspensions, where the gel and the glass transitions interplay,
and investigate the behaviour of the dynamical susceptibility for wavector ranging from 
$k=0.1$ to $6.28$.  We show that $\chi_4(k,t)$ has a complex behaviour, 
but that the analysis of the dependence on the wave vector allows to isolate the gel-like features and the 
glassy-like ones. The gel-like behaviour dominates at small wave vectors, the glassy-like behaviour at large ones, and 
finally, the combined effect of both transitions at intermediate ones, in analogy  with results for models of colloidal gel \cite{jstat2009}.
We also discuss the dependence of the self Intermediate Scattering Function, on the wave vectors, and 
the connection with static structure of the system.
 
\section{Methods}
We consider the same model investigated in Ref. \cite{nagi}: 
a $50{:}50$ binary mixture of $N = 10^3$ hard spheres (monomers) of mass $m$ and diameters $\sigma$ and 
$1.4\sigma$,
in a box of size $L$ with periodic boundary conditions.
The volume fraction $\phi = Nv/L^3$, where $v$ is the average particle volume,
is tuned by changing the size $L$ of the box. The mass $m$, the diameter $\sigma$ of the smaller particles,
and the temperature $T$, fix mass, length and energy scales, while the time unit is  $\sqrt{m\sigma^2/T}$.

The model is studied using event driven molecular dynamics simulations~\cite{lu91,alti87}.
After thermal equilibration at the desired volume
fraction, permanent bonds are introduced with probability $p$
between any pair of particles separated by less than $1.5\sigma$.
A bond corresponds to an infinite square well potential, extending from
$\sigma$ to $1.5\sigma$.
The procedure used to insert the bonds mimics
a light--induced polymerization process,
as the number of bonds depends on both $p$ and $\phi$.
Here, we consider bond probability $0.4$ and  different values of the volume fraction.
For each set of parameters, we simulate 
$30-50$ realisations of the system with different bond configurations.
In Ref. \cite{berthier} a similar model, in which the bond lifetime may be suitably modulated, has been extensively studied.

\section{Results}
\subsection{Phase diagram}
\begin{figure}[ht]
\begin{center}
\resizebox{0.6\columnwidth}{!}{%
  \includegraphics{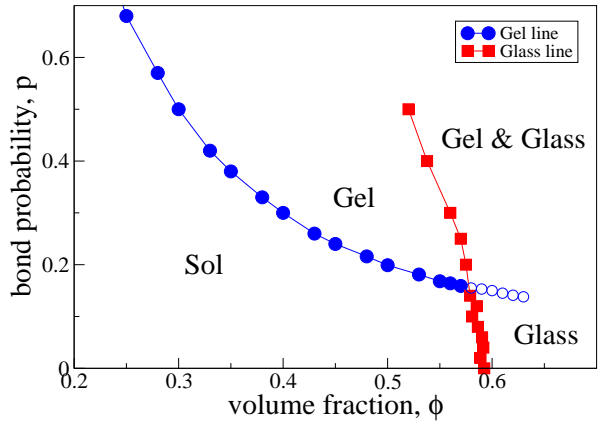} }
\end{center}
\caption{Structural arrest diagram as a function of the volume fraction,
$\phi$, and of the bonding probability, $p$, illustrating the
interplay of the gel and the glass transition lines. The gel line
is determined via percolative analysis, after introducing bonds
with probability $p$ in an equilibrium (full symbols) or in an
out--of--equilibrium (open symbols) monomer suspension. The glass
line is defined as that where the extrapolated diffusion
coefficient vanishes.
Solid lines are guides to the eye (from Ref. \cite{nagi}).}
\label{fig:1}       
\end{figure}

We start by reviewing the phase diagram of the investigated model, 
where the gel is characterised by the presence of a percolating
cluster, and  the sol-gel transition is identified with the percolation line \cite{nagi}, 
following Ref.s \cite{flory,degennes}.
A standard finite--size scaling analysis of the mean
cluster size~\cite{stauffer} is applied to identify the percolation line, $p_{gel}(\phi)$, i.e. the dependence of the critical
value of the bond probability on the volume fraction. As illustrated in Fig.\ref{fig:1},
it is found that $p_{gel}(\phi)$ decreases as $\phi$ increases.

The mean squared displacement is evaluated, and it has been observed that the diffusion coefficient at given bond probability, $p$, 
decreases as a power law $|\phi-\phiglass|^c$, approaching a critical value of the volume fraction, $\phiglass(p)$, depending on $p$. 
This allowed to identify the glass transition line, $\phi_{glass}(p)$, also illustrated in Fig.\ref{fig:1}. 
Remarkably, the phase diagram shown in Fig.\ref{fig:1} is akin to that obtained in the MCT \cite{gotze3,gotze1,gotze2} $F_{13}$ model  
\cite{goetze}, with the gel line corresponding to the continuous transition line of the MCT \cite{arenzon} and
the glassy line to the discontinuous one. These two lines intersect, the
glassy line entering  in the gel region, where the liquid-glass transition becomes a gel-glass transition.
In the MCT $F_{13}$ model, the discontinuous transition ends on a high order critical point ($A_3$ singularity) \cite{gotze3,dawson}
characterised by logarithm decay of the relaxation functions. 
Due to long relaxation time involved, it is rather difficult to localise such
singularity. However, evidence of logarithmic decay is found \cite{nagi} in
a region inside the gel phase, close to the glass transition line.

\subsection{Self Intermediate Scattering Function and Dynamical Susceptibility}
In order to connect the static structure to the dynamical behaviour, we evaluate 
the self Intermediate Scattering Function (sISF), $F_s(k,t)$, and the dynamical susceptibility, $\chi_4(k,t)$, defined respectively as:
\begin{equation}
 F_s(k,t) = \left[ \langle \Phi_s(k,t) \rangle  \right],
\label{self}
\end{equation}
\begin{equation}
 \chi_4(k,t) = N \left[ \langle |\Phi_s(k,t)|^2 \rangle -\langle \Phi_s(k,t) \rangle ^2\right],
\label{chi4}
\end{equation}
where 
$\Phi_s(k,t)=\frac{1}{N}\sum_{i=1}^N e^{i\vec{k}\cdot(\vec{r}_i(t)-\vec{r}_i(0))}$, $\langle\dots\rangle$ is the thermal average, 
$\left[\dots\right]$ is the average over the bond configurations, and the sums are done on all  particles. 
\begin{figure}[ht]
\begin{center}
\resizebox{1.0\columnwidth}{!}{%
 \includegraphics{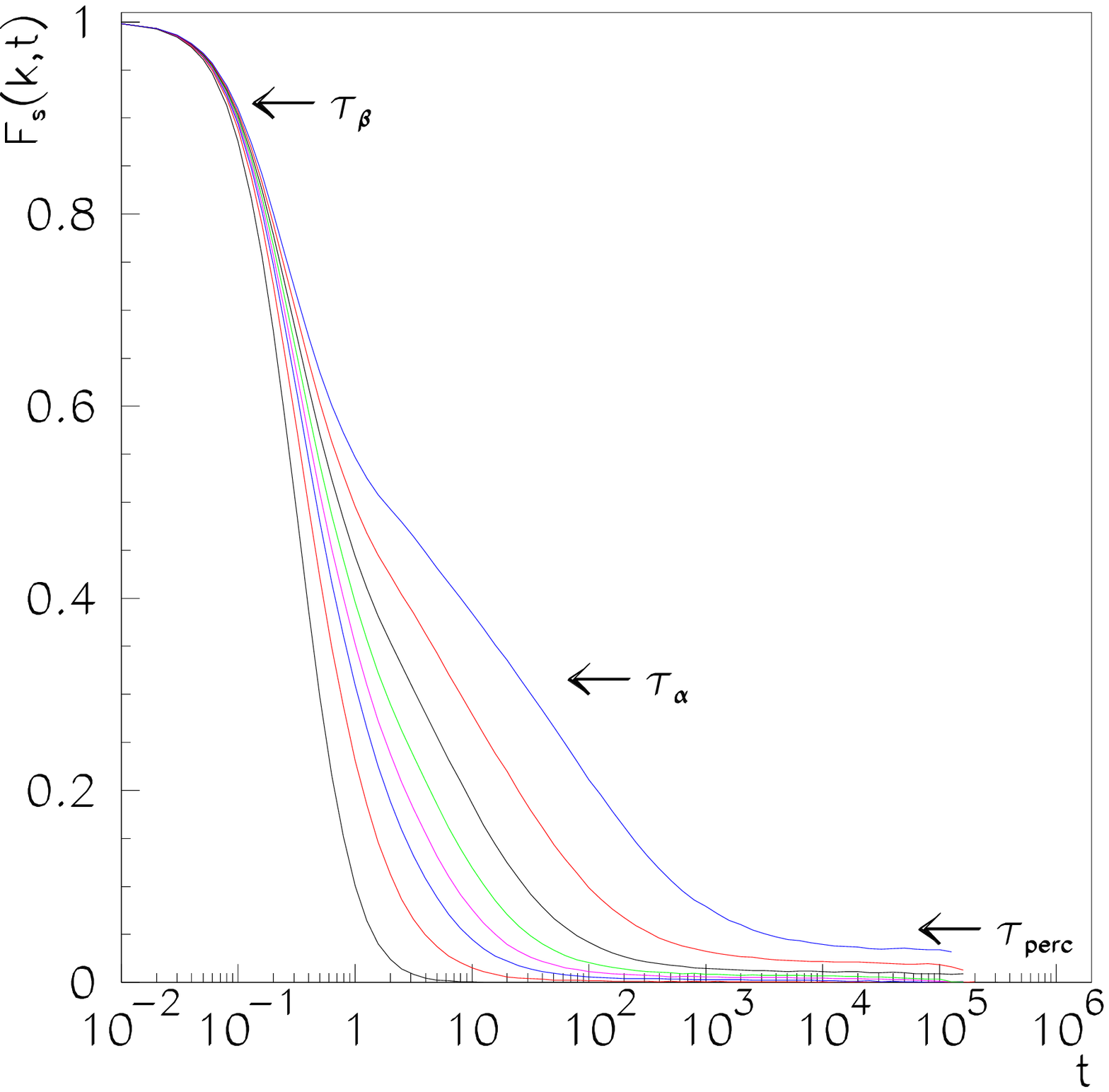}
 \includegraphics{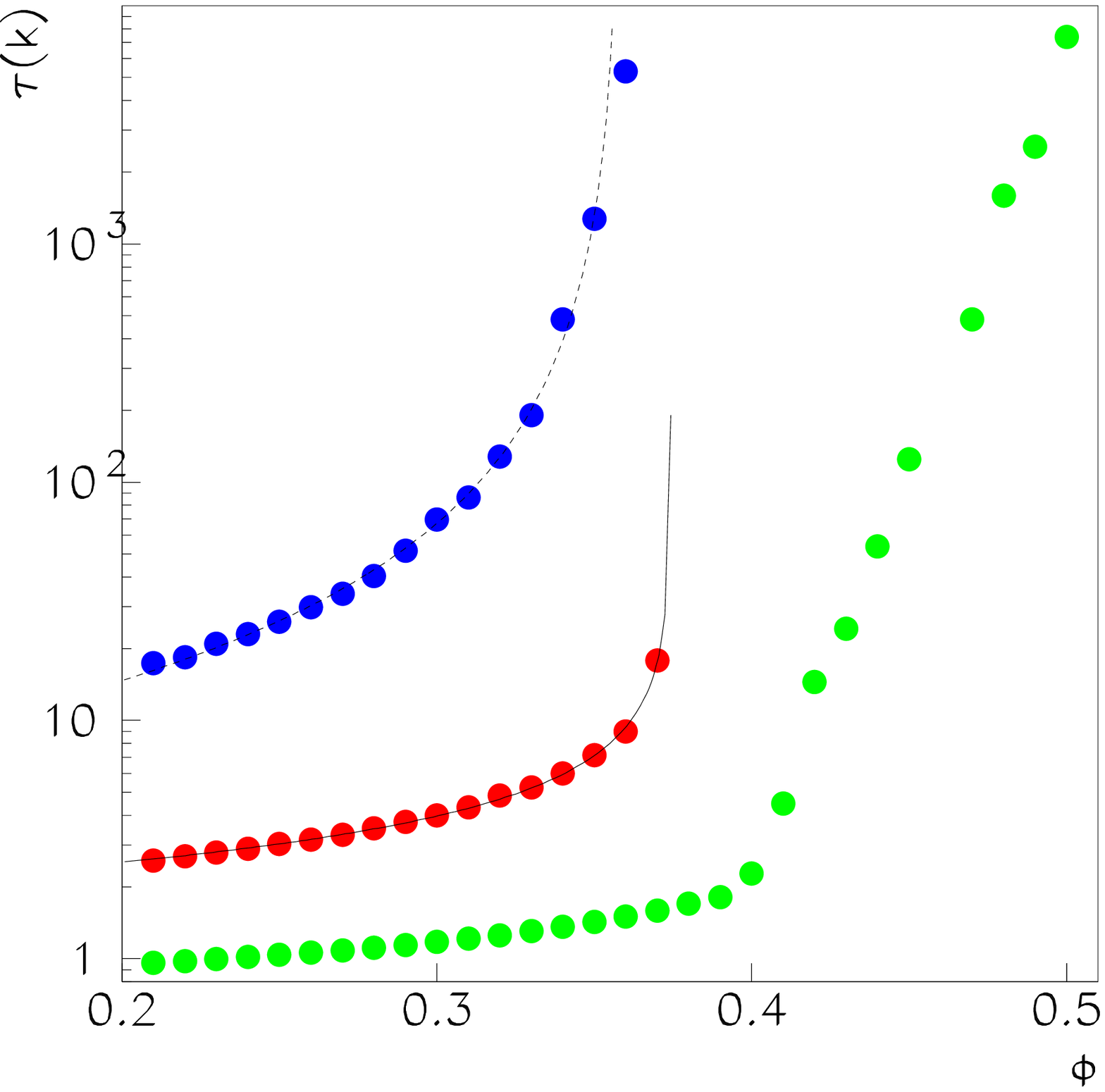}
}
\end{center}
\caption{a) Self ISF, $F_s(k,t)$ for $p=0.4$ and $k=6.28$ at different volume fraction $\phi=0.4$, $0.45$, $0.47$, $0.48$, $0.49$, $0.5$, $0.51$, 
$0.52$ (from left to right).
b) Relaxation time, $\tau(k)$, for $p=0.4$ and $k=1$ (blue circles), $k=3$ (red circles), $6.28$ (green circles),
as function of the volume fraction $\phi$.
Lines in figure are power law fitting functions,
$A (\phi_f-\phi)^{-\gamma}$, with
and $A=0.89$, $\phi_f=0.36$ and $\gamma=1.52$ (dotted line) for $k=1$,
and $A=1.02$, $\phi_f=0.37$ and $\gamma=0.52$ (continuous line) for $k=3$.
}
\label{fig:2}
\end{figure}

Three different relaxation time scales are recognised \cite{nagi}
in the relaxation functions: $\tau_\beta$, due to the rattling of particles in cage formed by the neighbors \cite{review_jump,JSTAT_jump,SM_corr}; 
$\tau_\alpha>\tau_\beta$, due to the opening
of the cage, diverging at the glass transition line; and finally
$\tau_{perc}>\tau_\alpha$, due to the relaxation of the largest cluster, diverging at
the gel transition line (see Fig. \ref{fig:2}a). Similar findings are obtained in Ref. \cite{berthier}.
At $\phi>\phi_{gel}$, $F_s (k,t)$, does not relax to zero, and reaches at long time a
finite value, due to particles belonging to the spanning cluster, that decreases as the wave vector $k$ increases, and increases as a function of 
the volume fraction.
Thus, evaluating the integral relaxation time, 
$\tau(k)\equiv\int dt~ t~ F_s(k,t)/\int dt~ F_s(k,t)$, we expect $\tau(k)$ to diverge at the percolation transition

In Fig.\ref{fig:2}b, $\tau(k)$ is plotted as function of the volume fraction for different wave vectors. 
We observe a divergence of  $\tau(k)$ for $k=1$ and $k=3$ roughly at the percolation threshold, 
$\phi_c\sim 0.37$, obtained from the divergence of the mean cluster size,  
whereas $\tau(k)$ for $k=6.28$ smoothly increases for $\phi>\phi_{gel}$.
We suggest that the behaviour of the relaxation time, at $k=6.28$,
is a numerical artifact: at large wave vectors and small volume fraction,
due to the numerical accuracy reachable in the simulations, 
the plateau, reached at long time, is so small that it is not observable at all
in our data (see Fig. \ref{fig:2}a), and apparently it seems that the relaxation time is finite also in the gel phase.

So, the percolating cluster dominates the self ISF long time decay. Conversely, as it has been also observed in Ref. \cite{nagi}, finite clusters 
dominate the long time mean square displacement, which results to be diffusive also in the gel phase, with a diffusion 
coefficient vanishing only at the glass transition line. 

\begin{figure}
\begin{center}
\resizebox{1.0\columnwidth}{!}{%
  \includegraphics{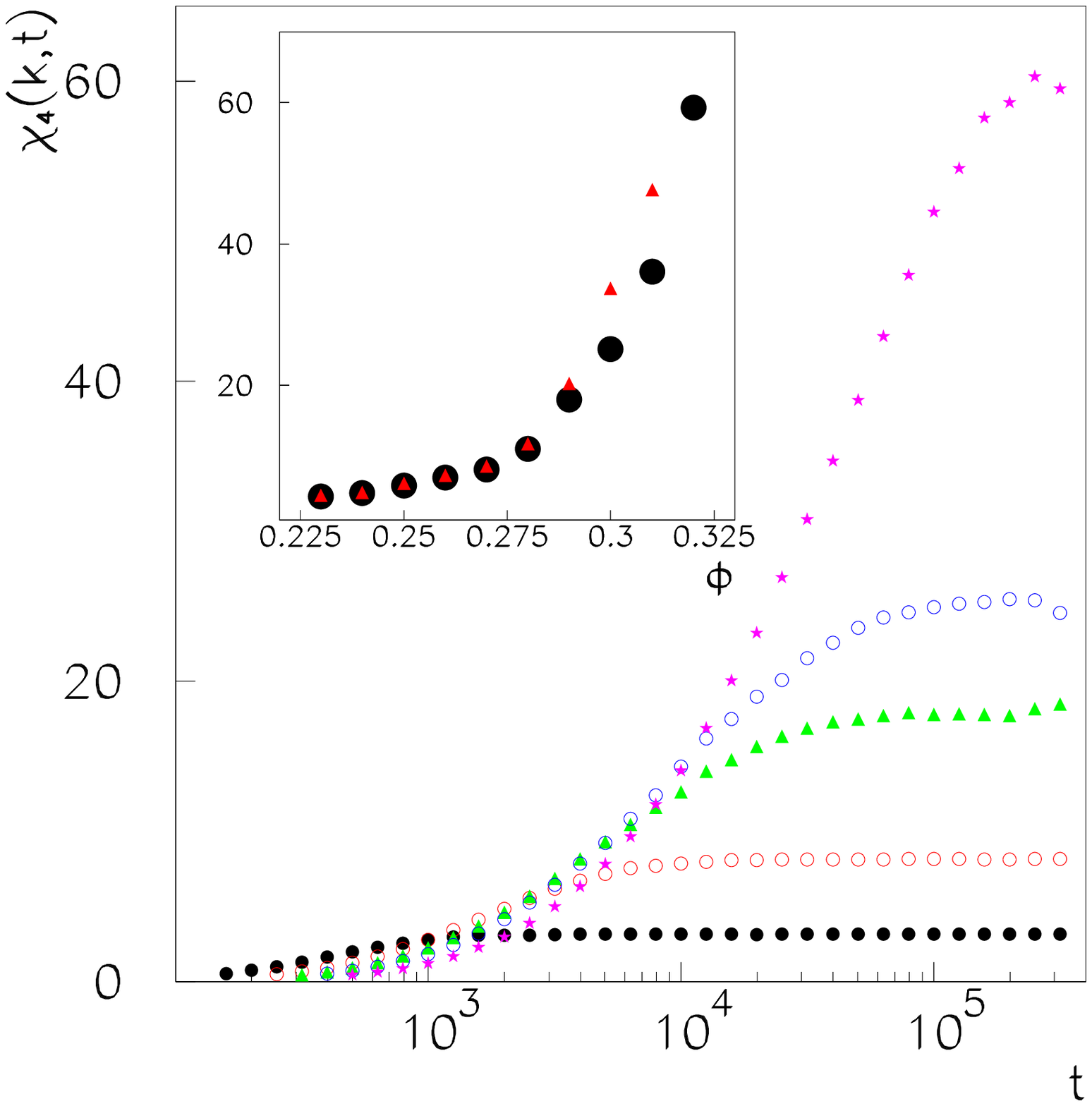} 
  \includegraphics{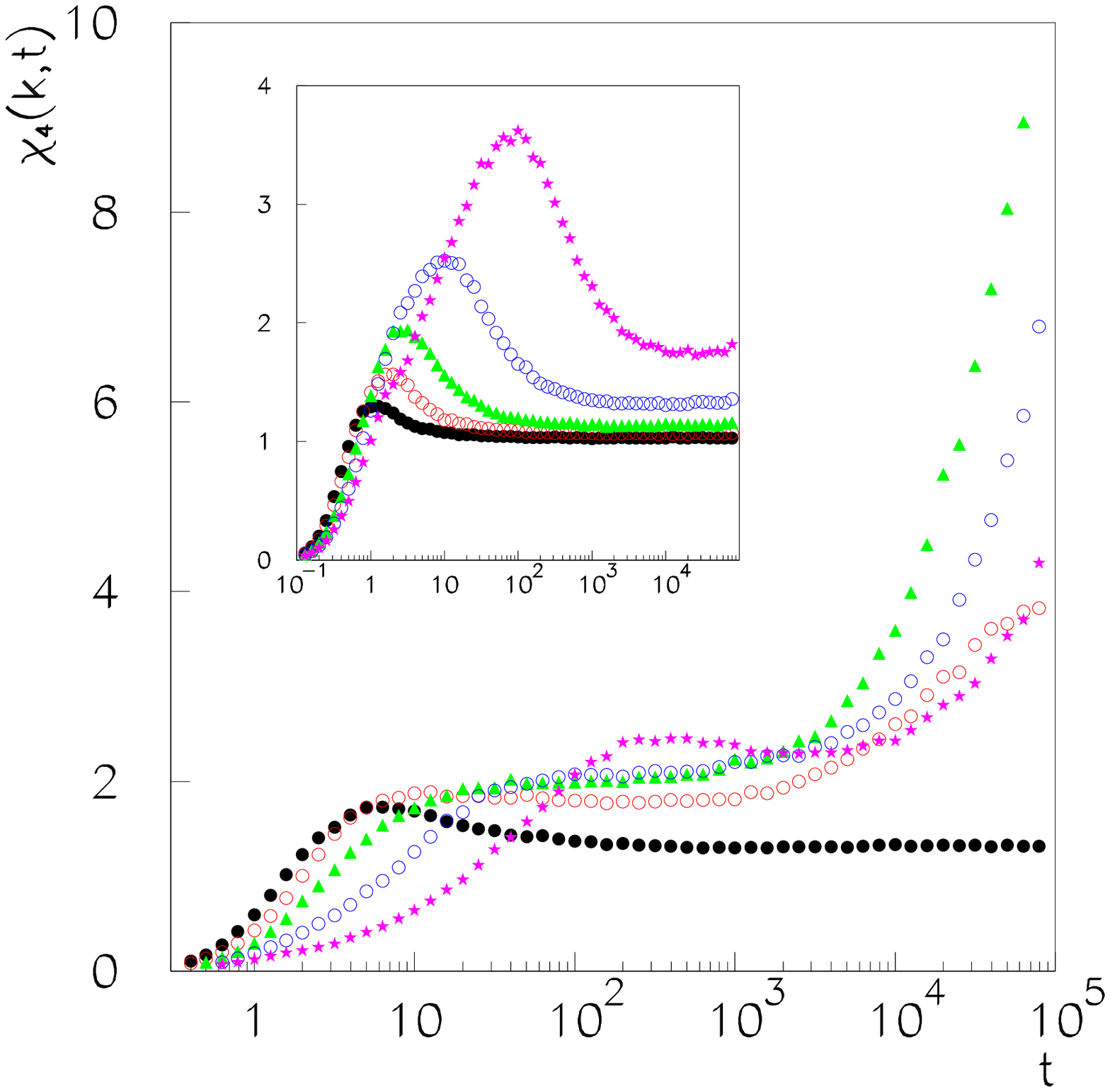} }
\end{center}
\caption{
a) {\bf Main frame}: Dynamical susceptibility, $\chi_4(k,t)$, for $p=0.4$, $k=0.1$ and $\phi=0.21$, $0.27$, $0.29$, $0.3$, $0.32$ (from bottom to top).
{\bf Inset}: Asymptotic value of the dynamical susceptibility (black circles) and mean cluster size of the large particles (red triangles).
b) {\bf Main frame}: Dynamical susceptibility, $\chi_4(k,t)$, for $p=0.4$, $k=3$ and $\phi=0.4$, $0.44$, $0.47$, $0.5$, $0.52$ (same symbols as in the 
inset).
{\bf Inset}: Dynamical susceptibility $\chi_4(k,t)$ for $p=0.4$, $k=6.28$ and $\phi=0.4$, $0.44$, $0.47$, $0.5$, $0.52$ (from bottom to top).}
\label{fig:3}
\end{figure}

In Ref. \cite{abete}, the gel formation is studied in a model system undergoing a chemical gelation by means of molecular dynamics simulations. 
Approaching the gelation threshold from the sol phase, the dynamic susceptibility is found to be a
monotonic function increasing with time, which tends in the limit of long times to a plateau,
whose value diverges, as a function of the distance from the gelation transition, as the mean cluster size.
Moreover, it has been theoretically shown that, in general in chemical gels, in the thermodynamics limit, 
at small enough wave vectors $k$, such as $2\pi/k > \xi$ with $\xi$
the average linear size of the largest cluster, the dynamic susceptibility obtained from the self ISF actually tends to the mean cluster size. 
In the following, we will check this prediction in the model here studied.
In Fig. \ref{fig:3}a, the dynamical susceptibility, $\chi_4(k,t)$, is plotted for $k=0.1$.
Indeed, we observe that $\chi_4(k,t)$
tends to a plateau (see main frame of Fig.\ref{fig:3}a) that coincides with the mean cluster size at small volume fraction (see inset of Fig. 
\ref{fig:3}a). A deviation is observed approaching the percolation threshold, where $\xi$ diverges and the condition $2\pi/k > \xi$ does not hold 
yet.
At the largest wave vector ($k=6.28$), due to the crowding of particles, $\chi_4(k.t)$  displays a maximum, as usually observed 
in glassy systems, whose value increases as the glass transition is approached (see inset of Fig.\ref{fig:3}b),
and, due to the presence of permanent clusters, a plateau different from $1$ at long times. 
At intermediate wave vector ($k=3$), the superposition of these two mechanisms gives origin to more complex features, as we can see in main frame of
Fig.\ref{fig:3}b.

\section{Discussion}
The presence of permanent bonds in gelling system generates correlation between the positions of pairs of particles belonging to the same 
cluster, which manifests as a plateau in the dynamical susceptibility.
Interestingly this type of behaviour is very similar to that observed
in a spin glass model \cite{lattice-gas}, and it is the signal that heterogeneities in chemical gel have a static nature.
In the limit of small wave vector, this plateau coincides with the mean cluster size and diverges at the gelation threshold.
In systems, where both glass and gel transitions are present, a second mechanism, due to the crowding, contributes to the dynamical susceptibility, 
as correlation between the displacements of different particles.
Here, we have clarified that these two mechanisms can be unentangled considering the wave vector dependence of the dynamical susceptibility.
The superposition of these two mechanisms originates a range of different behaviours, depending on the wave vector.

Finally we would like to dedicate this paper to Professor Alberto Robledo
for his great scientific achievements on the occasion of his $70^{th}$
Birthday.

\section*{Acknowledgements}
We acknowledge financial support from MIUR-FIRB
RBFR081IUK, from the SPIN SEED 2014 project
{\it Charge separation and charge transport in hybrid
solar cells}, and from the CNR-NTU joint laboratory {\it Amorphous materials for energy harvesting
applications}.


\begin{thebibliography}{0}
\bibitem{flory}
P.J. Flory {\it Principles of Polymer Chemistry}, Cornell University Press (Ithaca, NY, 1954).

\bibitem{degennes} 
P.G. de Gennes {\it Scaling Concepts in Polymer Physics}, Cornell University Press (Ithaca, NY, 1993).

\bibitem{jcp}
A. Fierro, T. Abete, A. Coniglio, The Journal of chemical physics {\bf 131} 194906 (2009).

\bibitem{gotze3}
W. G\"otze, J. Phys. Cond. Matt. {\bf 11}, A1 (1999).

\bibitem{arenzon} J.J. Arenzon, A. Coniglio, A. Fierro, M. Sellitto,
 Phys. Rev. E {\bf 90}, 020301(R) (2014);
A. Coniglio, J.J. Arenzon, A. Fierro, M. Sellitto M, {\it Eur. Phys. J. Special Topics} {\bf 223}, 2297 (2014).

\bibitem{gotze1}
W. G\"otze {\it Complex dynamics of glass-forming
liquids}, (Oxford University Press, Oxford, 2009).

\bibitem{gotze2}
W. G\"otze, L. Sjogren, Rep. Prog. Phys. {\bf 55}, 241 (1992)

\bibitem{JSTAT_DH}
R. Pastore, A. de Candia, A. Fierro, M. Pica Ciamarra, A. Coniglio J. Stat. Mech. 074011 (2016).


\bibitem{abete}
T. Abete, A. de Candia, E. Del Gado, A. Fierro, A. Coniglio, Phys. Rev. Lett. {\bf 98}, 088301 (2007);
T. Abete, A. de Candia, E. Del Gado, A. Fierro, A. Coniglio, Phys. Rev. E {\bf 78},  041404 (2008).


\bibitem{het1}
W. Kob, C. Donati, S.J. Plimpton, P.H. Poole and  S.C. Glotzer, Phys. Rev. Lett. {\bf 79}, 2827 (1997).
 
\bibitem{het2}
S. Franz,  G. Parisi, J. Phys.: Condens. Matter {\bf 12}, 6335 (2000). 

\bibitem{het3}
L. Berthier et al., Science {\bf 310}, 1797 (2005).
 
\bibitem{het4}
G. Biroli, J.P. Bouchaud , K. Miyazaki, D.R. Reichman, Phys. Rev. Lett. {\bf 97}, 195701 (2006).
 
\bibitem{het5}
L. Berthieret al., J. Chem. Phys. {\bf 126}, 184503 (2007)
\bibitem{het7}
C. Dalle-Ferrier et al., Phys. Rev. E {\bf 76}, 041510 (2007).
 
\bibitem{het8}
R. Pastore, M. Pica Ciamarra, A. Coniglio, Fractals {\bf 21}, 1350021 (2013).


\bibitem{chen}
F. Mallamace, S.H. Chen, A. Coniglio, L. de Arcangelis, E. Del
Gado, A. Fierro, Phys. Rev.  E {\bf 73}, 020402 (2006);
S.H. Chen, W.R. Chen, F. Mallamace, Science {\bf 300}, 619 (2003);
F. Mallamace, C. Corsaro,  H.E. Stanley, D. Mallamace, S. H. Chen, J. Chem. Phys. {\bf 139}, 214502 (2013).

\bibitem{berthier} P. Chaudhuri, P.I Hurtado, L. Berthier, W. Kob, J. Chem. Phys {\bf 142}, 174503 (2015);
P. Chaudhuri, L. Berthier, P.I. Hurtado, W. Kob, Phys. Rev. E {\bf 81}, 040502 (2010).

\bibitem{nagi}
N. Khalil, A. de Candia, A. Fierro, M. Pica Ciamarra, A. Coniglio, Soft Matter {\bf 10}, 4800 (2014).

\bibitem{jstat2009} A. de Candia,E.  Del Gado, A. Fierro, A. Coniglio, J. Stat. Mech. P02052 (2009).

\bibitem{lu91} B.D. Lubachevsky, Journal of Computational Physics {\bf 94}, 255 (1991).
 
\bibitem{alti87} M.P. Allen, D.J. Tildesley, {\it Computer Simulation of Liquids}  (Oxford University Press, Oxford, 1987).
 
\bibitem{stauffer}
D. Stauffer, A. Aharony, {\it Introduction to percolation theory} (Taylor $\&$ Francis, London, 1992).

\bibitem{goetze}
W. G\"otze, M. Sperl Phys. Rev. E {\bf 66}, 011405 (2002).

\bibitem{dawson} 
K. Dawson,  M. Foffi, M. Fuchs, W. G\"otze , F. Sciortino, M. Sperl, P. Tartaglia, T. Voigtmann, E. Zaccarelli, Phys. Rev. E {\bf 63}, 011401 (2000).

\bibitem{review_jump} M. Pica Ciamarra, R. Pastore, A. Coniglio, Soft Matter {\bf 12}, 358 (2016).

\bibitem{JSTAT_jump} R. Pastore, A. de Candia,  A. Fierro, A. Coniglio, M. Pica Ciamarra, {\it J. Stat. Mech.} 054050 (2016).

\bibitem{SM_corr} R. Pastore, A. Coniglio, M. Pica Ciamarra, Soft Matter {\bf 11}, 7214 (2015).

\bibitem{lattice-gas} A. Fierro, A. de Candia, A. Coniglio,  Phys. Rev. E {\bf 62}, 7715 (2000).


\end{thebibliography}
\end{document}